\documentclass[12pt]{iopart}
\usepackage{amssymb}
\usepackage{amsopn}
\usepackage{graphics}
\usepackage{graphicx}
\usepackage{psfrag}
\newcommand{\ff}[2]{\frac{#1}{#2}}
\newcommand{\dd}[0]{\mathrm{d}}
\newcommand{\bb}[0]{\begin{eqnarray}}
\newcommand{\ee}[0]{\end{eqnarray}}
\newcommand{\nn}{\nonumber}
\newcommand{\fd}{\frac{1}{2}}

\newcommand{\PF}[0]{\mathcal{Z}}
\newcommand{\pf}[0]{\mathcal{Q}}
\newcommand{\prodd}[2]{\overrightarrow{\prod_{#1}^{#2}}}
\newcommand{\prodg}[2]{\overleftarrow{\prod_{#1}^{#2}}}

\newcommand{\bQp}{\mathbb{Q}^{\mathrm{p}}}
\newcommand{\bL}{\mathbb{L}}
\newcommand{\D}{\mathcal{D}}
\newcommand{\action}{\mathcal{S}}
\newcommand{\fZ}{Z_{q+\fd}(u)}
\newcommand{\fZzero}{Z_{q+\fd}(0)}
\newcommand{\dt}{\Delta t}
\newcommand{\tth}{\tilde\theta}
\newcommand{\sump}[1]{\mathop{\rm Tr}_{#1}}
\DeclareMathOperator*{\TR}{\mathfrak{Tr}}

\begin{document}
\title[]{1D action and partition function for the 2D Ising model with a
boundary magnetic field}
\author{Maxime Clusel$^\dag$ and  Jean-Yves Fortin$^\ddag$}
\address{\dag\ Laboratoire de Physique, \'Ecole normale sup\'erieure de Lyon, 
46 all\'ee d'Italie, 69364 Lyon Cedex 07, France}
\address{\ddag\ Laboratoire de Physique Th\'eorique, 
Universit\'e Louis Pasteur, 
3 rue de l'Universit\'e, 67084 Strasbourg Cedex, France}
\ead{maxime.clusel@ens-lyon.fr, fortin@lpt1.u-strasbg.fr}

\begin{abstract}
In this article we present an alternative method to that developed by 
B. McCoy and T.T. Wu to obtain some exact results for the 2D Ising model
with a general boundary magnetic field and for a finite size system.
This method is a generalisation of ideas from V.N. Plechko presented for 
the 2D Ising model in zero field, based on the representation of the 
Ising model using a Grassmann algebra. A Gaussian 1D action is
obtained for a general configuration of the boundary magnetic
field. When the magnetic field is homogeneous, we
check that our results are in agreement with McCoy and Wu's previous
work. This 1D action is used to compute in an efficient way 
the free energy in the special case of an inhomogeneous boundary magnetic 
field. This method is useful to obtain new exact results for interesting 
boundary problems, such as wetting transitions.
\end{abstract}
 \submitto{\JPA}
\maketitle
%
\section{Introduction}
Studied for the first time in 1925 \cite{ising}, the Ising model is
one of the most important models of statistical physics. The two
dimensional case, solved exactly for the first time in 1944 by Onsager
\cite{onsager}, is the prototype of systems with second order phase
transition and non Gaussian critical exponents. It has therefore been
studied extensively by various exact and approximate methods. 
One important result is when Schultz, Mattis and Lieb \cite{SML,LSM,LM}
used transfer matrix method, Jordan-Wigner transformation and fermionization 
to simplify the Onsager solution in zero magnetic field, 
whereas fruitful links with 1D Quantum Field Theory and Conformal Field 
Theory have been developed in a more recent period to provide a more 
general frame for studying 2D critical systems. 
In particular, a way to express the Ising Hamiltonian as a Gaussian 
Grassmannian action was established long time ago \cite{samuel,nojima}, 
and this idea was extended by Plechko 
to compute the partition function of the 2D Ising model in zero field for a
large class of lattices \cite{plechko85,plechko88}, using operator
symmetries that simplify the algebra of transfer matrix. \\
The 2D Ising model has also been a starting point to study some
boundary problems, for example the effect of a boundary magnetic
field (BMF) on the propagation of a domain wall which is similar 
to a wetting or pinning problems. 
It is quite remarkable that exact results can be
obtained in this case since the model with a general uniform field
has not yet been solved except at the critical
temperature \cite{zamolo1,zamolo2} where conserved quantities have been
found. In a series of papers, McCoy and Wu computed the partition function 
associated with a uniform BMF on one side of a square lattice and evaluated 
the boundary correlation functions, using dimer 
statistics\cite{mccoybook,wu66,mccoy67,cheng67,mccoy67b,mccoy68}. 
 With transfer matrix and spinor methods, some exact results have also been 
found for configurations with different fixed boundary spins or equivalently 
infinite BMF\cite{abraham80,abraham82,abraham88}, or finite BMF
\cite{yangfisher75}. These methods use mainly
Jordan-Wigner transformation and spinor method
related to rotation matrices\cite{spinormethod1,spinormethod2}.
One interesting case is a configuration with infinite BMF on the two sides
of an infinite strip with opposite sign\cite{abraham80}. 
A single wall domain pinned along the middle of the strip is present at 
low temperature, separating two regions of opposite magnetisation. 
Diffusive or sharp interfaces occur depending on temperature. 
Another case is when each boundary is composed of spins up followed at
one point by spins down alongside of an infinite strip
\cite{interfaceAbraham82,interfaceAbraham84}. 
From this point, an interface develops perpendicular to the strip inside 
the bulk.
Other possible configurations with fixed boundary spins were studied
using the same methods \cite{abraham88,GrainBoundary}.
A solution with finite BMF was also proposed using Boundary Quantum 
Field Theory \cite{chatterjee}, or Conformal Field Theory in continuous
systems\cite{mussardoBMF}.\\ 
Our aim here is to present an alternative derivation of the exact partition 
function including a general BMF, extending the method
introduced by Plechko. This method appears to be simpler than the
McCoy and Wu's derivation, or transfer matrix methods, and allows exact 
calculations for more complicated cases of BMF with finite amplitude. 
By this we demonstrate that the partition function can be simply expressed 
as a Grassmann path 
integral of a 1D Gaussian action with general or random magnetic fields after
integrating over the bulk degrees of freedom. In the special case of
an homogeneous BMF, we are able to compute the
free energy on the lattice, and boundary spin-spin correlation
functions. We check that our formula is equivalent to the one obtained
by McCoy and Wu \cite{mccoybook,mccoy67b}. Moreover, we also apply
this method to obtain the exact partition function and the free energy 
when there is an interface developing between two finite opposite fields 
$h$ and $-h$ on one boundary side, as the simplest application of an 
inhomogeneous BMF.
To our knowledge, this case has not been considered in the literature,
except for infinite fields \cite{abraham80}, 
and is very similar to a problem of an interface pinning on a boundary with 
the strength of the pinning that can be tuned (here the magnitude of the 
field).
This method can be easily extended for more general configurations and 
therefore is useful for instance to studying wetting 
problems \cite{abraham80,ebner90}.
\\
The article is organised as follow: In section 2, we introduce the
notations used throughout the article. In section 3 and 4, following
Plechko's method, we obtain a Grassmann path integral representation of the
partition function and the main action. In section 5, as a
preliminary and useful exercise, we solve in the same way the 1D Ising model 
with an homogeneous magnetic field, in order to introduce the method to
the reader. Section 6 is dedicated to the explicit calculation of the
partition function for the two dimensional Ising model; we then give
the corresponding boundary 1D action after integrating over the bulk
variables. At this stage, we compare our
results with those of McCoy and Wu, and Au-Yang and Fisher, 
taking the thermodynamic limit. In the
section 7, we compute the two-point correlation function on the
boundary, which is necessary, in section 8, to obtain the expressions of
the partition function and the free energy in the case of a pinned interface
on one boundary with an inhomogeneous BMF.

%
%
\section{General notation}
\begin{figure}
\begin{center}
\includegraphics[scale=0.6]{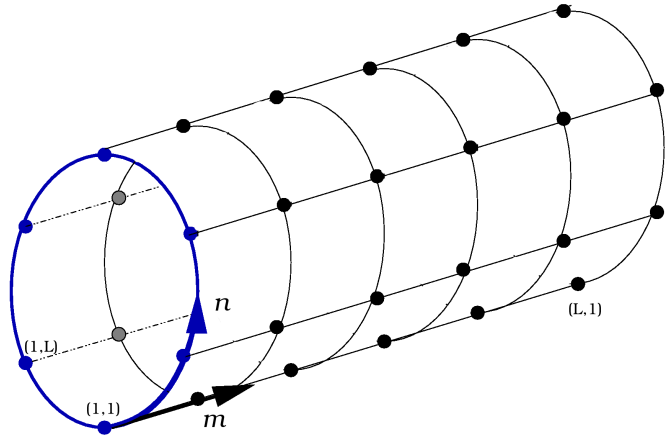}
\caption{Description of the model on a periodical lattice with free
conditions in one direction.}
\label{lattice}
\end{center}
\end{figure}

In the following we consider the Ising model on a square lattice of
size $L$ with spins $\sigma_{mn}=\pm 1$. For sake of simplicity, we
limit ourself here to the case where the coupling constant $J$ is the same in
both directions. The method works however if there are two different
coupling constants along vertical and horizontal bounds. Until section
4 we consider inhomogeneous magnetic fields $h_n$, placed on the sites
of the first column $m=1$  (see Figure \ref{lattice}).  Periodic
boundary conditions for the spins are imposed in the direction
parallel to the magnetic field line, $\sigma_{m1}=\sigma_{mL+1}$,  and
free boundary conditions in the transverse direction, formally
equivalent to $\sigma_{0n}= \sigma_{L+1n}=0$.\\
The Hamiltonian is then given by \bb
\mathcal{H}=-J\sum_{m,n=1}^{L}(\sigma_{m n}\sigma_{m+1 n}+ \sigma_{m
n}\sigma_{m n+1})-\sum_{n=1}^{L}h_n\sigma_{1n}.  \ee The partition
function (PF) $\PF$ is defined as
\[\PF=\sum_{\{\sigma_{mn} \} } \exp(-\beta \mathcal{H}),\]
where $\beta=1/k_B T$ and the sum is over all possible spin
configurations.  We can write, using $\sigma_{mn}^2=1$,
\bb \fl \nn e^{-\beta \mathcal{H}}=\left[\cosh(\beta
J)\right]^{2L^2}\prod_{n=1}^{L} \cosh(\beta h_n)(1+u\sigma_{1n})\cdot
\prod_{m,n=1}^{L}(1+t\sigma_{mn}\sigma_{m+1n})
(1+t\sigma_{mn}\sigma_{mn+1}), \ee with $u_n=\tanh(\beta h_n)$ and
$t=\tanh(\beta J)$. We then define $\pf$ as
\bb
\label{defpf}
\fl
\pf[h]=\sump{\sigma_{mn}}\left[ \prod_{m,n=1}^{L}(1+t\sigma_{mn}
\sigma_{m+1n})(1+t\sigma_{mn}\sigma_{mn+1})\cdot\prod_{n=1}^{L}(1+u_n\sigma_{mn})
\right], \ee where $\sump{\sigma}$ is the normalised sum $\frac{1}{2}
\sum_{\sigma=\pm 1}$.  This PF has already been calculated by McCoy
and Wu \cite{mccoybook} for a uniform field on a boundary. This was done in two
steps. First they proved that this PF  is the Pfaffian of a matrix,
using dimer statistics, they then performed the direct calculation of
this Pfaffian. Here we treat the problem in a different way. Our idea
is to generalise the elegant method  introduced by Plechko
\cite{plechko85,plechko88} for the 2D Ising model in zero field since this method appears to be simple, and provides  a direct
link with Quantum Field Theory. It is indeed straightforward
to obtain the expression of the quadratic fermionic action. Here we
show that we can derive such a quadratic action in presence of a
general  boundary magnetic field.
%
%
\section{Transformation of the PF using a Grassmann representation}
Following Plechko, we introduce pairs of Grassmann variables in order
to remove the local interaction between spins. We briefly define some
useful tools using Grassmann algebra. For more details, we refer the
reader to the book by Nakahara \cite{nakahara}.\\ A Grassmann algebra
$\mathcal{A}$ of size $N$  is a set of $N$ anti-commuting objects
$\{a_i\}_{i=1,N}$ satisfying:
$$ \forall \ 1\le i,j\le N,\ a_ia_j=-a_ja_i,$$
which implies $a_i^2=0$. Functions defined on such an algebra are
particularly simple, they are polynomials. It is possible to define a
notion of integration \cite{nakahara} with the following rules:
\bb \int \dd a\; a=1, \int \dd a\; 1=0, \ee and for any function
$f(a)$,
\bb \int \dd a \;f(a)=\ff{\partial f(a)}{\partial a}.  \ee With these
definitions, Gaussian integrals are expressed by
\bb \int \prod_{i=1}^{N}\dd a_i^* \dd a_i \exp\left(\sum_{i,j=1}^{N}
a_iM_{ij} a^*_j\right)=\det M.  \ee We also define a trace operator
over an algebra $\mathcal{A}=\{a,a^*\}$ as
\bb \TR_{a,a^*}\left[f(a,a^*)\right] \equiv \int \dd a^* \dd a \;
f(a,a^*)e^{aa^*}, \ee with the simple rules
\[\TR_{a,a^*}\left[1\right]=1,\;\;\TR_{a,a^*}\left[aa^*\right]=1.\]
This operator will be useful in the following, and the subscripts may
be omitted implicitely 
when the trace is performed over the Grassmann  variables
that are present in the expressions inside the brackets.  Grassmann
variables are introduced in the PF in order to decouple the  spin
variables. Terms containing the same spin are then put together and
the sum over the spin configurations is performed. We use the fact that
\bb \fl \label{link} 1+t\sigma\sigma'=\int \dd a^*\dd a(1+a\sigma)
(1+ta^*\sigma')e^{aa^*}=\TR [(1+a\sigma)(1+ta^*\sigma')] \ee and
follow closely the notation in reference \cite{plechko85}.  We
consider the following link variables
\bb \eqalign{
\label{psi1}\psi_{mn}^{(1)}=1+t\sigma_{mn}\sigma_{m+1n},\\
\label{psi2}\psi_{mn}^{(2)}=1+t\sigma_{mn}\sigma_{mn+1}.
} \ee In order to decouple the products of two spins, we can express
each object as a trace operator over a product of two Grassmann
polynomials using equation (\ref{link})
\bb \eqalign{ \psi^{(1)}_{mn}=\TR\left[A_{mn}A^*_{m+1n}\right],\\
\psi^{(2)}_{mn}=\TR\left[B_{mn}B^*_{mn+1}\right], } \ee where \bb
\eqalign{
\label{monome}
A_{mn}=1+a_{mn}\sigma_{mn},\; A^*_{mn}=1+ta^*_{m-1n}\sigma_{mn}, \\
B_{mn}=1+b_{mn}\sigma_{mn},\; B^*_{mn}=1+tb^*_{mn-1}\sigma_{mn}.  } \ee
The next step is to gather the different terms corresponding to the
same spin. For that we consider the mirror symmetries introduced by
Plechko \cite{plechko88} and the operations of moving Grassmannian
objects. Such operations are possible only within the trace operator. 
For example, the functions $\psi_{mn}^{(1)}$ and $\psi_{mn}^{(2)}$
are real functions, and therefore commute with each other. We can
also write $\psi_{mn}^{(1)}\psi_{mn}^{(2)}=\TR\left [
(A_{mn}A^*_{m+1n})(B_{mn}B^*_{mn+1})\right ]$. Inside the brackets $[\dots ]$,
the two groups $(A_{mn}A^*_{m+1n})$ and $(B_{mn}B^*_{mn+1})$ can be 
moved independently since we can perform the trace on each of them
separately, which gives real and therefore commuting quantities like
$\psi_{mn}^{(1)}$ or $\psi_{mn}^{(2)}$ that can be moved and inserted
everywhere, and, at the end of these operations, we reintroduce the 
different integrations or trace function. For example we have 
\bb 
\fl \psi_{mn}^{(1)}\psi_{mn}^{(2)}&=\psi_{mn}^{(2)}\psi_{mn}^{(1)},\\
\nn &=\TR\left [
(A_{mn}A^*_{m+1n})(B_{mn}B^*_{mn+1})\right ]
=\TR\left [
(B_{mn}B^*_{mn+1})(A_{mn}A^*_{m+1n})\right ],\\
\nn &=\TR\left [(B_{mn}(A_{mn}A^*_{m+1n})B^*_{mn+1})\right ]
=\TR\left [
(A_{mn}(B_{mn}B^*_{mn+1})A^*_{m+1n})\right ],
\ee
where the brackets $(\dots)$ define the commuting group of Grassmannian
objects.
In general, if we have three groups of commuting objects 
$\{(\mathcal{O}_i\mathcal{O}^*_i)\}_{i=1..3}$ we easily obtain the 
following mirror symmetries that can be applied to the objects
(\ref{monome}):
\bb\label{mirror} 
({\cal O}_1^*{\cal O}_1)({\cal O}_2^*{\cal O}_2)({\cal
O}_3^*{\cal O}_3) &=({\cal O}_1^*({\cal O}_2^*({\cal O}_3^* {\cal O}_3){\cal
O}_2){\cal O}_1)
\\ \nn &=({\cal O}_3^*({\cal O}_2^*({\cal O}_1^* {\cal O}_1){\cal
O}_2){\cal O}_3).  
\ee 
It is also important to
treat the spin boundary conditions separately  from the bulk
quantities in order to obtain an expression valid not only in the
thermodynamic limit, but for all finite values of $N$. Indeed, in the
direction parallel to the line of magnetic fields where
$\sigma_{mL+1}=\sigma_{m1}$, the corresponding link element can be
expressed as \bb
\label{boundary}
\psi^{(2)}_{mL}=\TR\left[B_{mL}B^*_{mL+1}\right]
=\TR\left[B^*_{m1}B_{mL}\right], \ee 
where
$B^*_{m1}=1+tb^*_{m0}\sigma_{m1}$. The equality (\ref{boundary})
associated with definitions (\ref{monome}) imposes
$b^*_{m0}=-b^*_{mL}$. The periodic conditions on spins therefore lead
to anti-periodic conditions on Grassmann variables.  In the transverse
direction, we have $\sigma_{0n}=\sigma_{L+1n}=0$,  corresponding to
free boundary conditions. This implies the boundary conditions on Grassmann variables $a^*_{0n}=0$ and therefore
$A^*_{1n}=1$.  The PF $\mathcal{Q}[h]$ (\ref{defpf}) can be written in
terms of the  $\psi^{(k)}_{mn}$ as
\bb
\label{pfu}\pf[h]=\sump{\sigma_{mn}}\left[ \prod_{m,n=1}^{L}\psi^{(1)}_{mn}
\psi^{(2)}_{mn}\cdot\prod_{n=1}^{L}(1+u_n\sigma_{mn}) \right].  \ee
Using the mirror symmetry (\ref{mirror}), the boundary terms can be
written as
\bb \eqalign{ \prod_{n=1}^{L}
\psi_{Ln}^{(1)}=\TR\left[\prodd{n=1}{L}A_{Ln}\right],\\
\prod_{m=1}^{L} \psi_{mL}^{(2)}=\TR\left[\prodd{m=1}{L}B^*_{m1} \cdot
\prodg{m=1}{L}B_{mL}\right],  } \ee 
where we introduce the notation
\bb
\nn \prodd{k=1}{N}A_k=A_1\cdot A_2 \cdots A_k,\\
\nn \prodg{k=1}{N}A_k=A_N \cdot A_{N-1} \cdots A_1.
\ee
These products can be reorganised
as follow \cite{plechko88}:
\bb \prod_{n=1}^L \psi_{Ln}^{(1)}\cdot \prod_{m=1}^L \psi_{mL}^{(2)}
=\TR\left[\prodd{m=1}{L}B^*_{m1}\cdot \prodd{n=1}{L}A_{Ln}\cdot
\prodg{m=1}{L}B_{mL}\right].  \ee
For the bulk elements, we obtain the
following arrangement
\bb \prod_{m=1}^L\psi_{mn}^{(2)}=\TR \left[\prodg{m=1}{L} B_{mn} \cdot
\prodd{m=1}{L}B^*_{mn+1}\right],\\
\prod_{n=1}^{L-1}\prod_{m=1}^L\psi_{mn}^{(2)}\cdot
\prod_{m=1}^{L}\psi_{mL}^{(2)}= \TR\left[\prodd{m=1}{L}B^*_{m1}\cdot
\prod_{n=1}^{L-1}\prod_{m=1}^L\psi_{mn}^{(2)} \cdot
\prodg{m=1}{L}B_{mL}\right], \\ 
\label{expr}
\prod_{n=1}^{L}
\prod_{m=1}^{L}\psi_{mn}^{(2)} \cdot \prod_{n=1}^L
\psi_{Ln}^{(1)}=\TR\left[\prodd{n=1}{L}\left(
\prodd{m=1}{L}B^*_{mn}\cdot A_{Ln}\cdot \prodg{m=1}{L}B_{mn}
\right)\right], \ee
where we use the fact that $\psi_{Ln}^{(1)}$ are
commuting objects as well as the product reorganisation:
\bb\label{mirror2} 
{\cal O}_1^*({\cal O}_1{\cal O}_2^*)({\cal O}_2{\cal O}_3^*){\cal O}_3
=
\prodd{m=1}{3}{\cal O}_m^*{\cal O}_m.
\ee 
We now insert the
product over the remaining $\psi_{mn}^{(1)}$ inside the previous
expression (\ref{expr}) 
\bb \fl \TR \left[\prod_{m=1}^{L-1} \psi_{mn}^{(1)}  \cdot
\prodd{m=1}{L}B^*_{mn}\cdot A_{Ln}\ldots \right] = \TR
\left[B^*_{1n}A_{1n}\cdot\prodd{m=2}{L}A^*_{mn}B^*_{mn}A_{mn}\ldots
\right],\; \; \; \; \; \; \; \; \ee
and finally obtain
\bb \fl \prod_{m,n=1}^L\psi^{(1)}_{mn}\psi^{(2)}_{mn}=\TR\left[
\prodd{n=1}{L}B^*_{1n}A_{1n}  \left(\prodd{m=2}{L}A^*_{mn}B^*_{mn}A_{mn} \cdot \prodg{m=2}{L}B_{mn}\right) B_{1n}\right].\ee
The PF is rewritten in a way that the sum
on each spin can be performed by iteration
\bb
\label{repmix}
\fl \pf[h]=\sump{\sigma_{mn}} \TR\left[
\prodd{n=1}{L}B^*_{1n}A_{1n}(1+u_n\sigma_{1n})  \left(
\prodd{m=2}{L}A^*_{mn}B^*_{mn}A_{mn}\cdot \prodg{m=2}{L}B_{mn}
\right) B_{1n} \right].\; \; \; \; \; \; \; \; \ee In fact, as we do
not yet fermionize $(1+u_n\sigma_{1n})$, we reproduce here the
Plechko's derivation in the  special case of free-periodic boundary
conditions \cite{plechko85}. This expression is the basis of the rest
of this paper. The sum over the spins outside the magnetic field region
will lead to a quadratic action over Grassmann variables which
therefore commute with the rest of the elements belonging to the
first column $(1,n)$.

\section{Grassmannian representation of the action on the lattice}
The key point in equation (\ref{repmix}) is that the trace over the spin
configurations is performed in an iterative way. For example, the first
summation is done on spins $\sigma_{Ln}$ in the products
$A^*_{Ln}B^*_{Ln}A_{Ln}B_{Ln}$. This operation leads to a quantity
which is quadratic in Grassmann variables, which can be put outside
the general product (\ref{repmix}).  The same operation is then
performed on spins $\sigma_{L-1n}$ and so on.  This makes Plechko's
method efficient for the 2D Ising model in zero field. With a uniform
magnetic field in the bulk, the spin trace over the product of the
four previous operators would lead to a quantity which is linear and
quadratic in Grassmann variables and does not commute with the other
products.  However a BMF affects only the last products
depending on spins  $\sigma_{1n}$ and this makes the problem very
\textit{similar to a 1D Ising model in a uniform magnetic field}: It
is therefore solvable.

\subsection{Trace over spins inside the bulk}

For spins $\sigma_{mn}$ inside the bulk, $1\le n \le L$ and $2 \le m \le L$, 
we have to evaluate successively
\bb  \nn
\sump{\sigma_{mn}}\left[A^*_{mn}B^*_{mn}A_{mn}B_{mn}\right] &=
\exp(\mathbb{Q}_{mn}), 
\ee
with
\bb
\fl
\mathbb{Q}_{mn}&=a_{mn}b_{mn}+t^2a^*_{m-1n}b^*_{mn-1}+t(a^*_{m-1n}+b^*_{mn-1})(a_{mn}+b_{mn}).
\ee
\noindent These terms commute with all Grassmannian terms and can be
pulled out of the  remaining products. We obtain
\bb \sump{\{\sigma_{mn}\}_{m=2..L}} \left[
\prodd{m=2}{L}A^{*}_{mn}B^*_{mn}A_{mn}\cdot \prodg{m=2}{L}B_{mn}
\right]= \exp\left(\sum_{m=2}^{L}\mathbb{Q}_{mn}\right).  \ee
The PF can now be written as
\bb
\label{PFmix2}\pf[h]=\TR\left[
\exp\left(\sum_{n=1}^{L}\sum_{m=2}^{L}\mathbb{Q}_{mn}\right) \cdot
\prodd{n=1}{L}\underbrace{\sump{\sigma_{1n}}
\left((1+u_n\sigma_{1n})B^*_{1n}A_{1n}B_{1n}\right)}_{\mbox{
\footnotesize boundary spins $\sigma_{1n}$}}\right].  \ee

\subsection{Trace over the boundary spins}

In the expression (\ref{PFmix2}), we can evaluate separately the trace
over the spins $\sigma_{1n}$, leading to
\bb
\fl
\nn
\sump{\sigma_{1n}}[(1+u_n\sigma_{1n})B^*_{1n}A_{1n}B_{1n}]
=1+a_{1n}b_{1n}+tb^*_{1n-1}(a_{1n}+b_{1n})+u_n\mathbb{L}_{n}+
u_ntb^*_{1n-1}a_{1n}b_{1n},\\
\label{Trb}
\mathbb{L}_n=a_{1n}+b_{1n}+tb^*_{1n-1}.  \ee
\noindent The presence of a magnetic field on a site introduces a
linear Grassmann term.  This term no longer commutes with the others,
and we need to compute the  product in (\ref{PFmix2}) carefully. \\ We
would like to change artificially the fixed boundary conditions to periodic 
ones in order to simplify the subsequent calculations based on Fourier 
transformation.  
The quadratic part of equation (\ref{Trb}) is equal to $\mathbb{Q}_{1n}$  
with fixed boundary conditions: We can write
\bb
\mathbb{Q}_{1n}=\mathbb{Q}^{\mathrm{p}}_{1n}-ta^*_{Ln}\mathbb{L}_{n},
\\ \nn
\mathbb{Q}_{1n}^{\mathrm{p}}=a_{1n}b_{1n}+t^2a^*_{0n}b^*_{1n-1}+
t(a^*_{0n}+b^*_{1n-1})(a_{1n}+b_{1n}), \ee where we introduce the
boundary quantities $ a_{0n}^*=a_{Ln}^*$.
$\mathbb{Q}^{\mathrm{p}}_{1n}$ corresponds to periodic boundary
conditions for the Grassmann variables (or anti-periodic conditions
for  the spins).  We obtain \bb
\sump{\sigma_{1n}}\left[(1+u_n\sigma_{1n})B^*_{1n}A_{1n}B_{1n}\right]
=\exp\left(\mathbb{Q}^{\mathrm{p}}_{1n}+u_n\mathbb{L}_n-ta^*_{Ln}
\mathbb{L}_n\right).  \ee The correction to periodic conditions due to
the free boundary conditions for the spins is included in
$-ta^*_{Ln}\mathbb{L}_n$.
\subsection{Grassmann variables associated with the magnetic field}
Here we introduce a pair of Grassmann variables $(\tilde h_n,\tilde h^*_n)$
associated with the
BMF. In the rest of the article, we will refer to
it as the fermionic magnetic field. We have
\bb\label{Grassmannmagneticfield} 
\fl
\exp(u_n\bL_n)=1+u_n\bL_n=\int \dd
\tilde h_n^* \dd \tilde h_n\; (1+u_n\tilde h_n)
(1+\tilde h_n^*\bL_n) e^{\tilde h_n \tilde h_n^*}, \ee therefore
\bb
\nn
\fl
\sump{\sigma_{1n}}\left[(1+u_n\sigma_{1n})B^*_{1n}A_{1n}B_{1n}\right]
=\TR_{\tilde h_n,\tilde h_n^*}\left[ \exp\left(\mathbb{Q}^{\mathrm{p}}_{1n}+
(\tilde h_n^*-ta_{Ln}^*)\bL_n+u_n\tilde h_n\right)\right]. \ee 
We now perform a translation in the fermionic magnetic field
\bb
H_n=\tilde h_n,\;\;H^*_n=\tilde h^*_n-ta_{Ln}^*,
\ee
which leads to
\bb\nn
\label{trbord}
\fl \prodd{n=1}{L}\sump{\sigma_{1n}}
\left[(1+u_n\sigma_{1n})B^*_{1n}A_{1n}B_{1n}\right]
=\TR_{H_n,H_n^*}\left[\exp\left(\sum_{n=1}^L \bQp_{1n}+H^*_n\bL_n+
H_nta_{Ln}^*\right)\cdot \prodd{n=1}{L}e^{u_nH_n}\right].  
\ee 
It is
also useful to write the last $L$ products as a non local action along
the boundary line 
\bb
\label{prodexp}
\prodd{n=1}{L}e^{u_nH_n}=\exp\left(\sum_{n=1}^Lu_nH_n+\sum_{m=1}^{L-1}\sum_{n=m+1}^Lu_mu_nH_mH_n\right).  
\ee

\subsection{Fermionic action of the PF}
Putting equation (\ref{prodexp}) into (\ref{PFmix2}), we
obtain the Grassmannian representation of the PF
\bb
\label{Grep} \pf[h]=\int \D a^* \D a\:\D b^*\D b\: \D H^*\D H \;
\exp\mathcal{S}[a,a^*,b,b^*,H,H^*], \ee with the action $\mathcal{S}$
defined as  \bb
\label{action} 
\eqalign{ \mathcal{S}=&\sum_{mn=1}^{L}(\bQp_{mn}
+a_{mn}a^*_{mn}+b_{mn}b^*_{mn})
+\sum_{n=1}^{L}H^*_n\bL_n+\sum_{n=1}^{L}H_nta_{Ln}^*\\
&+\sum_{m<n}u_mu_nH_mH_n+\sum_{n=1}^{L}H_nH^*_n.  } \ee This action
can be separated into three terms  \bb
\action=\action_{\mathrm{bulk}}+
\action_{\mathrm{int}}+\action_{\mathrm{field}}, \ee with  \bb
\eqalign{
\action_{\mathrm{bulk}}&=\sum_{mn=1}^{L}(\bQp_{mn}+a_{mn}a^*_{mn}
+b_{mn}b^*_{mn}),\\
\action_{\mathrm{field}}&=\sum_{m<n}u_mu_nH_mH_n+\sum_{n=1}^{L}H_nH^*_n,\\
\action_{\mathrm{int}}&=\sum_{n=1}^{L}H^*_n\bL_n+\sum_{n=1}^{L}H_nta_{Ln}^*.
} \ee
The PF written as (\ref{Grep}) is just a Gaussian integral over the
set of variables $(a,a^*,b,b^*,H,H^*)$. If we first integrate over the
variables  $(a,a^*,b,b^*)$ corresponding to the action
$\action_{\mathrm{bulk}}+\action_{\mathrm{int}}$, we obtain a new
Gaussian action depending only on fermionic magnetic field $(H_n,H^*_n)$. This
new action is very similar to that for a one dimensional problem.
Actually the way we integrate (\ref{Grep}) is close to solving a  1D
Ising model with a magnetic field and Grassmann variables. In the next
section we present briefly this case since we will use similar tools
later. Our method can then be checked using the transfer matrix
techniques.

\section{1D Ising model with a homogeneous magnetic field}\label{1Dsection}
The treatment is similar to the 2D Ising model, except that there is
only one kind of link variables and no mirror symmetry involved. The
exact  solution in the case of a homogeneous magnetic field
$u=\tanh(\beta h)$ using the transfer matrix method is simply
\bb\label{transf1D} \fl 2^L\pf_{1\mathrm{D}}(h)=\left
(1+t+\sqrt{(1-t)^2+4tu^2}\right)^L+\left
(1+t-\sqrt{(1-t)^2+4tu^2}\right)^L.  \ee If we apply the Grassmann
transformations as before,  we can write an equation similar to
(\ref{Grep}):
\bb \pf_{1\mathrm{D}}(h)=\int\D a^* \D a\:\D H^*\D H\;
\exp(\action_{\mathrm{bulk}}+\action_{\mathrm{int}}
+\action_{\mathrm{field}}), \ee with
\bb \nn
\action_{\mathrm{bulk}}=\sum_{n=1}^{L}(a_na_n^*+ta_{n-1}^*a_n),\;\;
\action_{\mathrm{int}}=\sum_{n}H_{n}^*\bL_n, \\ \nn
\action_{\mathrm{field}}=\sum_{n}H_nH_{n}^*+u^2\sum_{m<n}H_mH_n,\;\;
\bL_n=a_n+ta_{n-1}^*. \ee The Grassmann variables $a_n$ and $H_n$ are
anti-periodic in space,  and can be Fourier transformed using
$a_n=\frac{1}{\sqrt{L}}\sum_{q}r_{q+\fd}^na_{q+\fd}$ with
$r_q=e^{2i\pi q/L}$.  In the new basis, the actions are almost
diagonalized
\bb\label{action1DFourier} \nn \fl \action_{\mathrm{bulk}}+
\action_{\mathrm{int}}=\sum_{q=0}^{L/2-1}(1-tr_{q+\fd})a_{q+\fd}a^*_{q+\fd}
+(1-t\bar r_{q+\fd})a_{-q-\fd}a_{-q-\fd}^* \\
\lo{+}\sum_{q=0}^{L/2-1}H_{q+\fd}^*(a_{q+\fd}+t\bar
r_{q+\fd}a_{-q-\fd}^*) +
\sum_{q=0}^{L/2-1}H_{-q-\fd}^*(a_{-q-\fd}+tr_{q+\fd}a_{q+\fd}^*), \ee
where the bar defines the complex conjugate. For $L$ odd, we have to
take care of the momenta on the diagonal of the Brillouin zone: it
leads to additional terms in (\ref{action1DFourier}); however, these
terms are irrelevant for large $L$. In the following, we restrict
ourselves to $L$ even. \\ The non local sum in
$\action_{\mathrm{field}}$ can be written in the Fourier modes as
\bb
\sum_{m<n}H_mH_n&=\sum_{q}\frac{1}{r_{q+\fd}-1}H_{q+\fd}H_{-q-\fd}, \\
\nn &=\sum_{q=0}^{L/2-1}\left (\frac{1}{r_{q+\fd}-1}- \frac{1}{\bar
r_{q+\fd}-1}\right )H_{q+\fd}H_{-q-\fd}.  \ee We can separate the
previous sums (\ref{action1DFourier}) into independent blocks of four
Grassmann variables $(a_{q+\fd},a_{q+\fd}^*,a_{-q-\fd},a_{-q-\fd}^*)$.
To compute the individual block integrals, we use the general Gaussian
formula:
\bb \nn \label{formula1D}
\fl
\int \dd a^* \dd a \: \dd b^* \dd b \;\exp(\alpha
aa^*+\bar\alpha bb^*+va+v^*a^* +wb+w^*b^*) =\alpha\bar\alpha\exp\left (
-\frac{ww^*}{\bar\alpha}-\frac{vv^*}{\alpha}\right ),
\ee
where $(v,v^*,w,w^*)$ are Grassmann variables and ($\alpha,\bar\alpha$)  two
independent complex numbers. We obtain \bb\nn
\pf_{1\mathrm{D}}(h)=\pf_{1\mathrm{D}}(0)\int\D H^*\D H  \exp\left (\sum_{q=0}^{L/2-1}
H_{q+\fd}H_{q+\fd}^*+H_{-q-\fd}H_{-q-\fd}^*\right .  \\ \label{res0}
\left. +\gamma^{1\mathrm{D}}_{q+\fd}H_{-q-\fd}^*H_{q+\fd}^*+
u^2\delta_{q+\fd}H_{q+\fd}H_{-q-\fd} \right ), \ee
where
\bb\label{res1Ddef}
\pf_{1\mathrm{D}}(0)=\prod_{q=0}^{L/2-1}|1-tr_{q+\fd}|^2,\;\;
\gamma^{1\mathrm{D}}_{q+\fd}=\frac{2it\sin \theta_{q+\fd}} {1-2t\cos
\theta_{q+\fd}+t^2}, \\ \label{defdelta}
\delta_{q+\fd}=-i\;{\rm{\cot}}(\fd\theta_{q+\fd}),
\;\;\theta_{q+\fd}=\frac{2\pi}{L}\left(q+\fd\right).  \ee
$\pf_{1D}(0)$ is the PF in zero field, and is equal to $1+t^L$ in this
case.\\  The remaining integrals over $(H,H^*)$ are easy to evaluate,
if we use
\bb \int \dd a^* \dd a \: \dd b^* \dd b \; \exp(aa^*+bb^*+\alpha
a^*b^*+\beta ab)=1-\alpha\beta.  \ee After some algebra and
simplifications, we finally obtain
\bb\label{res1D} \pf_{1\mathrm{D}}(h)=\prod_{q=0}^{L/2-1}\left (
1+t^2-2t\cos\theta_{q+\fd}+4tu^2\cos^2(\fd\theta_{q+\fd})\right ).
\ee Results (\ref{transf1D}) and  (\ref{res1D}) are equivalent when
$L$ is even but are written  in a different way. This has been checked
numerically for finite size systems, and analytically in the
thermodynamic limit.\\  Using Grassmann variables to express the PF in
terms of products over Fourier modes is of course less efficient in
the 1D case. However in 2D, the method is very similar  and leads to a
final expression which is similar to (\ref{res1D})  and (\ref{res0}) as
seen in the next section.
\section{Explicit calculation of the PF}
In this section, we perform the Gaussian integration of the
Grassmannian form of $\pf$ (\ref{Grep})~: We start by reducing the
number of Grassmann variables per site, then we integrate over the
variables in the bulk, in order to obtain a \textit{1D action},
expressed in terms of the fermionic magnetic field. Finally,
this last integral is evaluated, leading to the PF on the lattice. In
order to validate our method, we check that our result is identical
that obtained by McCoy and Wu, in the thermodynamic limit.

\subsection{Reduction of number of Grassmann variables per site}
In the 2D case, we can easily integrate half of the Grassmann
variables, for example $(a_{mn},b_{mn})$, by using the identity
\bb \int \dd b \dd a\; e^{ab+aL+b\bar L}=e^{\bar L L}.  \ee Since the
original measure is $\dd b^*\dd b \dd a^* \dd a$, moving $\dd b$ to
the right of  $\dd a^*$ implies a minus sign. After integrating over
$(a,b)$ we  define $c_{mn}=a_{mn}^*$ and $c_{mn}^*=-b^*_{mn}$, which
removes the  minus signs from the new measure $\dd c^* \dd c$.  We
thus obtain the following actions
\bb \eqalign{ \action_{\mathrm{bulk}}= \sum_{m,n=1}^{L}
c_{mn}c^*_{mn}+
t(c_{mn}^*+c_{mn})(c_{m-1n}-c^*_{mn-1})-t^2c_{m-1n}c^*_{mn-1}, \\
\action_{\mathrm{int}}= \sum_{n=1}^{L}
tH_nc_{0n}+(tc_{1n-1}^*+c^*_{1n}+c_{1n})H_n^*, \\
\action_{\mathrm{field}}=\sum_{m<n}u_m u_n
H_mH_n+\sum_{n=1}^{L}H_nH^*_n.  } \ee Taking into account the
different boundary conditions, we Fourier transform these variables as
in the 1D case
\bb c_{mn}=\ff{1}{L}\sum_{p,q=0}^{L-1}r_p^mr_{q+\fd}^n c_{pq+\fd},\;\;
c^*_{mn}=\ff{1}{L}\sum_{p,q=0}^{L-1}\bar{r}_p^m\bar{r}_{q+\fd}^n
c^*_{pq+\fd}, \ee and compute the bulk action
\bb \eqalign{ \action_{\mathrm{bulk}}=\sum_{p,q=0}^{L-1} \left (
1-t\bar r_{p}-tr_{q+\fd}-t^2\bar r_pr_{q+\fd} \right )
c_{pq+\fd}c^*_{pq+\fd}+ \\
-t\bar{r}_pc_{pq+\fd}c_{L-pL-q-\fd}+tr_{q+\fd}c_{pq+\fd}^*c_{L-pL-q-\fd}^*.
} \ee
\begin{figure}
\begin{center}
\psfrag{q}{$q+1/2$}
\psfrag{p}{$p$}
\psfrag{(0,0)}{$(0,0)$}
\psfrag{(L,L-1)}{$(L,L-1)$}
\includegraphics[scale=0.8]{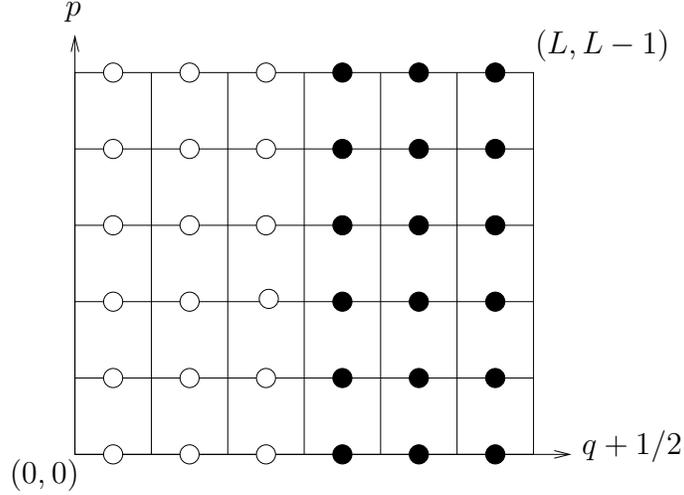}
\caption{Integration domain for the Fourier modes in the case where
$L$ is even ($L$=6). The set $S_1$ of modes corresponds to the white
points.  The other modes (set $S_2$) are obtained from set $S_1$ 
by the symmetry $(p,q) \rightarrow (L-p,L-q-1/2)$ modulo $L$.}
\label{fouriermode}
\end{center}
\end{figure}
The result implies a coupling between modes $(p,q+\fd)$ and
$(L-p,L-q-\fd)\sim(-p,-q-\fd)$. As in 1D, the sum can be expressed as
a sum over independent blocks containing the variables
$c_{pq+\fd},c_{-p-q-\fd},c_{pq+\fd}^*$ and $c_{-p-q-\fd}^*$.  These
different blocks are indeed independent  if we restrict to momenta
$(p,q)$ corresponding to the white points  (set $S_1$) of the
Brillouin zone in figure \ref{fouriermode}. 
In this case, the modes $(p,q+\fd)$  belonging
to $S_1$ and the modes $(L-p,L-q-\fd)$ (corresponding to $S_2$) completely
fill a Brillouin zone. This avoids counting the Grassmann variables
twice. We also need to make the action more symmetric, which can be written as
\bb \eqalign{ \fl \action_{\mathrm{bulk}}=\sum_{p,q\in S_1}\left (
\alpha_{pq+\fd}c_{pq+\fd}c^*_{pq+\fd}+\bar\alpha_{pq+\fd}
c_{-p-q-\fd}c^*_{-p-q-\fd}+  \beta_pc_{pq+\fd}c_{-p-q-\fd}\right.\\
\left.  \lo{+}\beta_{q+\fd}c_{pq+\fd}^*c_{-p-q-\fd}^* \right ),} \\
\alpha_{pq+\fd}=1-t\bar r_{p}-tr_{q+\fd}-t^2\bar r_pr_{q+\fd},\\
\beta_p=t(r_p-\bar r_p), \ee and for the interacting part
\bb \eqalign{ \action_{\mathrm{int}}= \frac{1}{L}\sum_{p,q\in S_1}
c_{pq+\fd}(r_pH^*_{q+\fd}-tH_{-q-\fd})+c^*_{pq+\fd}\bar{r}_p(1+tr_{q+\fd})H^*_{-q-\fd}+
\\
c_{-p-q-\fd}(\bar{r}_pH^*_{-q-\fd}-tH_{q+\fd})+c^*_{-p-q-\fd}r_p(1+t\bar{r}_{q+\fd})H^*_{q+\fd}.
} \ee The last action, $\action_{\mathrm{field}}$, is the same as in 1D.

\subsection{Integration over bulk variables : 1D action}
The integration over the variables $c$ and $c^*$ is performed using
the following identity, similar to formula (\ref{formula1D}):
\bb\nn \int \dd a^* \dd a\: \dd b^* \dd b \;\exp(\alpha
aa^*+\bar\alpha bb^*+ \beta ab+\bar \beta a^*b^*+av+bw+a^*v^*+b^*w^*)
\\ \label{formula2D} =(\alpha\bar\alpha-\beta\bar\beta) \exp\left [
\frac{1}{\alpha\bar\alpha-\beta\bar\beta} \left ( \bar\alpha
v^*v+\alpha w^*w+\bar\beta vw +\beta v^*w^*\right ) \right ].  \ee We
obtain
\bb\label{res2D} \fl\nn \int \D c^* \D c\;
e^{\action_{\mathrm{bulk}}+\action_{\mathrm{int}}}= \prod_{p,q\in S_1}
\left( \alpha_{pq+\fd}\bar{\alpha}_{pq+\fd}-\beta_p \beta_{q+\fd}
\right)\exp\left( \gamma_{pq+\fd} H^*_{-q-\fd} H^*_{q+\fd}+
\lambda_{pq+\fd}H_{q+\fd}H^*_{q+\fd}\right.\\ \left.
\lo{+}\bar{\lambda}_{pq+\fd}H_{-q-\fd}H^*_{-q-\fd}+
\epsilon_{pq+\fd}H_{-q-\fd}H_{q+\fd}\right), \ee 
where we have identified the different coefficients of the Grassmannian fields
forming a quadratic action:
\bb 
\nn \gamma_{pq+\fd}=\ff{1}{L}
\ff{1}{\alpha_{pq+\fd}\bar{\alpha}_{pq+\fd}-\beta_p \beta_{q+\fd}}
\left( -\alpha_{pq+\fd}(1+t\bar{r}_{q+\fd})
+\bar{\alpha}_{pq+\fd}(1+tr_{q+\fd})\right .  \\  \left
.+\beta_p(1+tr_{q+\fd})(1+t\bar{r}_{q+\fd}) +\bar{\beta}_{q+\fd}
\right),
\ee
and 
\bb\nn
 \lambda_{pq+\fd}=\ff{1}{L}
\ff{1}{\alpha_{pq+\fd}\bar{\alpha}_{pq+\fd}-\beta_p \beta_{q+\fd}}
\left(tr_p\beta_{q+\fd}+tr_p\alpha_{pq+\fd}(1+t\bar{r}_{q+\fd})
\right),\\ \epsilon_{pq+\fd}=\ff{1}{L}
\ff{t^2\beta_{q+\fd}}{\alpha_{pq+\fd}\bar{\alpha}_{pq+\fd}-\beta_p
\beta_{q+\fd}}.  \ee 
Inside the product (\ref{res2D}), the quantities 
in front of the exponentials can be
simplified using cosine functions,
\bb\nn \alpha_{pq+\fd}\bar{\alpha}_{pq+\fd}-\beta_p \beta_{q+\fd} =
(1+t^2)^2-2t(1-t^2)\left[\cos\theta_p+\cos\theta_{q+\fd}\right], \ee
and are invariant under the transformation $(p,q+\fd)\rightarrow
(L-p,L-q-\fd)$. In (\ref{res2D}), the product over $S_1$ of these 
coefficients define a bulk PF $\pf_0$:
\bb \pf_0^2= \prod_{p,q=0}^{L-1} \left[
(1+t^2)^2-2t(1-t^2)\left(\cos\theta_p+\cos\theta_{q+\fd}\right)
\right].  \ee 
In the thermodynamic limit and in zero field, the free energy per site
corresponding to $\pf_0$ is
equal to the one corresponding to $\pf[0]$, since the boundary conditions 
do not play any role on the bulk properties.
In this limit, the second order phase transition occurs
at a  temperature given by the solutions of the equation
$(1+t^2)^2-4t(1-t^2)=0$,  or $t_c=\sqrt{2}-1$, when the cosines, in
the long-wave length  limit $(p,q)\sim (0,0)$, approach unity. In this
case the free energy is singular.\\ The previous
coefficients $\gamma_{pq+\fd}$ are not symmetrical in  $(p,q)$ since
the model itself is not symmetrical in both directions.  However they
are antisymmetric $\gamma_{p,q+\fd}=-\gamma_{L-p,L-q-\fd}$. This is useful 
in order to reduce the summation over the variables
$(p,q)$ in the action (\ref{res2D}). This implies
\bb\label{res2D2} \nn \sum_{p,q\in S_1} \gamma_{pq+\fd} H^*_{-q-\fd}
H^*_{q+\fd}=
\sum_{q=0}^{L/2-1}\gamma_{q+\fd}^{2D}H^*_{-q-\fd}H^*_{q+\fd}, \ee with
\bb\nn \gamma^{2\mathrm{D}}_{q+\fd}=\fd \sum_{p=0}^{L-1} \left (
\gamma_{pq+\fd}-\gamma_{p-q-\fd} \right ). \ee
After simplification, we find that
\bb\nn \gamma_{pq+\fd}=\ff{1}{L}\ff{2it\sin\theta_{q+\fd}}
{(1+t^2)^2-2t(1-t^2)(\cos \theta_p+\cos\theta_{q+\fd})}.  \ee
and
\bb\label{defgamma}
\gamma_{q+\fd}^{2\mathrm{D}}=\ff{1}{L}\sum_{p=0}^{L-1}
\ff{2it\sin\theta_{q+\fd}}
{(1+t^2)^2-2t(1-t^2)(\cos\theta_p+\cos\theta_{q+\fd})}.  \ee 
The factors $\gamma^{2\mathrm{D}}$ play the role of Fourier coefficients
of an effective interaction between the boundary spins in the
magnetic field. We can notice from (\ref{res0}) that  coefficients
$\gamma^{1\mathrm{D}}$  describe the nearest neighbour interaction of
the 1D Ising model.  Here the spins on the boundary can be mapped onto
a 1D model.\\ The factors $\lambda_{pq+\fd}$ have a different
symmetry, $\lambda_{pq+\fd}=\bar{\lambda}_{-p-q-\fd}$, which allows
the same kind of manipulation as before.  Hence, defining  \bb
\eqalign{ \Lambda_{q+\fd}&=
\frac{1}{2}\sum_{p=0}^{L-1}(\lambda_{pq+\fd}+\bar{\lambda}_{p-q-\fd}),
\\ &=\frac{t}{L}\sum_{p=0}^{L-1}\frac{(1-t^2)\cos(\theta_p)-t(1+2t\cos
\theta_{q+\fd}+t^2)}{(1+t^2)^2-2t(1-t^2)(\cos\theta_p+\theta_{q+\fd})},
} \ee
we can write
\bb \nn \fl \sum_{p,q \in S_1} \lambda_{pq+\fd}H_{q+\fd}H^*_{q+\fd}+
\bar{\lambda}_{pq+\fd}H_{-q-\fd}H^*_{-q-\fd}=
\sum_{q=0}^{L/2-1}\Lambda_{q+\fd}H_{q+\fd}H^*_{q+\fd}+
\Lambda_{-q-\fd}H_{-q-\fd}H^*_{-q-\fd}.  \ee
Moreover, further reduction of the terms containing
$\epsilon_{pq+\fd}$ in the action (\ref{res2D}) leads to the following
simplification
\bb \nn \sum_{p,q \in S_1} \epsilon_{pq+\fd}H_{-q-\fd}H_{q+\fd}=
\sum_{q=0}^{L/2-1}t^2\gamma_{q+\fd}^{2\mathrm{D}}H_{-q-\fd}H_{q+\fd}.
\ee
Finally, the problem of the boundary field is reduced to a 
1D Gaussian action in Grassmann variables which is given by:
\bb
\label{res2D1Dgen}
\eqalign{ \fl \action_{\mathrm{1D}}= \sum_{q=0}^{L/2-1}\left[
(1+\Lambda_{q+\fd})H_{q+\fd}H^*_{q+\fd}+
(1+\Lambda_{-q-\fd})H_{-q-\fd}H^*_{-q-\fd}+
\gamma^{2\mathrm{D}}_{q+\fd} H^*_{-q-\fd}H^*_{q+\fd}\right.\\ \left.
\lo{-}t^2\gamma_{q+\fd}^{2\mathrm{D}}H_{q+\fd}H_{-q-\fd} \right] +
\sum_{q,q'=0}^{L-1}\Delta_{q,q'}[u]H_{q+\fd}H_{q'+\fd}, } \ee with \bb
\nn
\Delta_{q,q'}[u]=\frac{1}{L}\sum_{m<n}u_mu_nr_{q+\fd}^mr_{q'+\fd}^n.
\ee This action has the same form as that for the 1D problem
(\ref{action1DFourier}), except for the additional terms
$\Lambda$ and $-t^2\gamma^{2\mathrm{D}}$, which are not present
in the 1D case. This is due to the fact that $\pf_0$ is not the true
action in the zero field case: If we integrate (\ref{res2D1Dgen})
with respect to the Grassmann fields when $u=0$, this will lead to a
non zero corrective factor in front of $\pf_0$, which is however 
irrelevant in the thermodynamic limit (the free energy corresponding to this
factor is of order $L$ instead of $L^2$).  $\pf_0$ is
therefore not the finite size zero field partition function for the 
periodic/free
spin boundary case. The factor comes from the free
boundary conditions that restore these conditions.\\
Using this 1D Gaussian action, the partition function for the 2D Ising model 
with an inhomogeneous or random boundary magnetic fields reads :

\bb
\label{GenPF}
\pf[h]=\pf_0 \int\dd H^*\dd H
\exp \left ( \action_{1\mathrm{D}} \right ). 
\ee
In the following, we will compute the remaining Gaussian integrals in 
two special cases where $\Delta_{qq'}[u]$ simplifies.
The first one is the case of an homogeneous magnetic field, where
$\Delta_{qq'}[u]=u^2 \delta_{q+1/2} \delta(q+q')$ with $\delta(q-q')$ the
Kronecker symbol, and $\delta_{q+1/2}$ defined in
(\ref{defdelta}). The second case corresponds to the simplest case of
inhomogeneous magnetic field, when half of the boundary spins is subject to
$+H$ and the other half to $-H$.
\subsection{Expression for the partition function and thermodynamic limit}
In this section, we are interested in the special case of an
homogeneous BMF. The previous action
(\ref{res2D1Dgen}) reduces to: \bb
\label{res2D1D}
\nn \action_{\mathrm{1D}}= \sum_{q=0}^{L/2-1}\left[
(1+\Lambda_{q+\fd})H_{q+\fd}H^*_{q+\fd}+
(1+\Lambda_{-q-\fd})H_{-q-\fd}H^*_{-q-\fd}+ \right .  \\ \nn \left .
\gamma^{2\mathrm{D}}_{q+\fd} H^*_{-q-\fd}H^*_{q+\fd}-
t^2\gamma_{q+\fd}^{2\mathrm{D}}H_{q+\fd}H_{-q-\fd} \right]+
u^2\sum_{q=0}^{L-1}\delta_{q+\fd}H_{q+\fd}H_{-q-\fd}, \ee with
$\delta_{q+\fd}$ defined in (\ref{defdelta}).
The successive integrations over the blocks of Grassmann magnetic
fields are easy to perform and we obtain:

\bb\nn
\pf(h)=\pf_0 \prod_{q=0}^{L/2-1} \fZ,
\ee
with
\bb
\label{defZ}
\fZ \equiv
(1+\Lambda_{q+\fd})(1+\Lambda_{-q-\fd})+\gamma_{q+\fd}^{2\mathrm{D}}
(u^2\delta_{q+\fd}-t^2\gamma^{2\mathrm{D}}_{q+\fd}).  \ee
We factorize the previous expression, in order to distinguish between 
the boundary effect in zero field and the contribution due to the
magnetic field: 
\bb
\label{resfinal}
\nn \fl
\pf(h)=\pf_0\prod_{q=0}^{L/2-1}\left[(1+\Lambda_{q+\fd})(1+\Lambda_{-q-\fd})
-t^2(\gamma^{2\mathrm{D}}_{q+\fd})^2\right] \left[
1+\frac{u^2\delta_{q+\fd}\gamma^{2\mathrm{D}}_{q+\fd}}{
(1+\Lambda_{q+\fd})(1+\Lambda_{-q-\fd})
-t^2(\gamma^{2\mathrm{D}}_{q+\fd})^2} \right ].  \ee
The total free energy is therefore written as
\bb \label{Ftotal}
F(h)=-Lk_{\mathrm{B}}T \ln\cosh(\beta h)-k_{\mathrm{B}}T \ln \pf_0
+Lf_{\mathrm{b}}+Lf_{\mathrm{field}}, \ee
with $f_\mathrm{b}$ a corrective free energy that
is needed to restore the free boundary conditions in the direction
transverse to the field,
\bb \beta f_{\mathrm{b}}= -\frac{1}{L}\sum_{q=0}^{L/2-1}  \ln\left[
(1+\Lambda_{q+\fd})(1+\Lambda_{-q-\fd})
-t^2(\gamma^{2\mathrm{D}}_{q+\fd})^2\right], 
\ee
and $f_{\mathrm{field}}$ the magnetic contribution to the free energy
\bb
\label{free_field_h}
\beta f_{\mathrm{field}}= -\frac{1}{L}\sum_{q=0}^{L/2-1} \ln \left[
  1+\frac{u^2\delta_{q+\fd}\gamma^{2\mathrm{D}}_{q+\fd}}{
    (1+\Lambda_{q+\fd})(1+\Lambda_{-q-\fd})
    -t^2(\gamma^{2\mathrm{D}}_{q+\fd})^2} \right ].  \ee
\begin{figure}[!htp]
\psfrag{JCh/(kbT)}{$JC_h /k_B T$}
\psfrag{kbT/J}{$k_BT/J$}
\centering \includegraphics[scale=0.5]{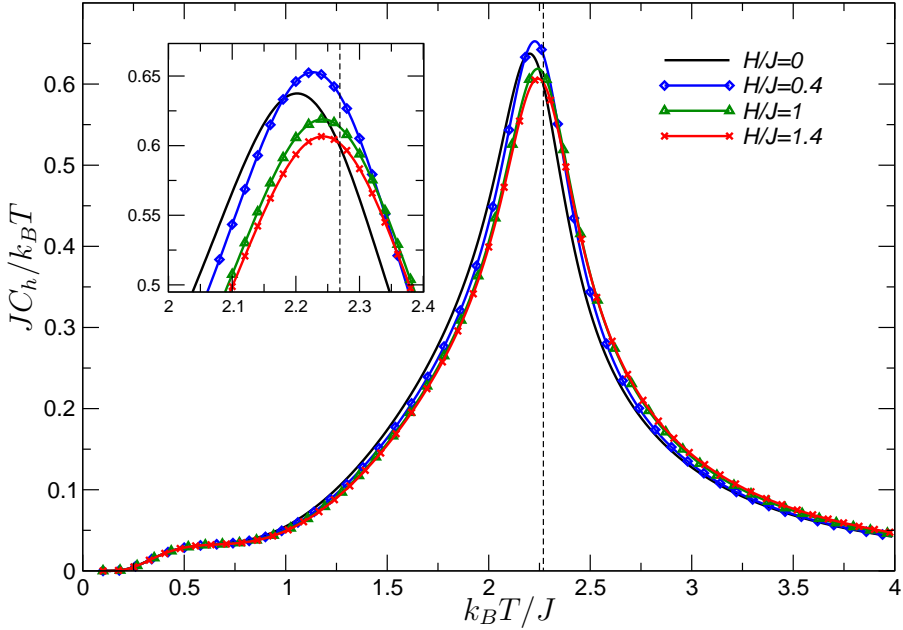}
\label{Free_vH}
\caption{Specific heat for various values of homogeneous magnetic field on the
boundary for $L$=20. The inset in a zoom in the region of the maximum. The vertical dashed line is the position of critical temperature for 2D Ising model in zero field, $k_B T_\mathrm{c}/J\simeq 2.26$. See also the same field dependence
in reference \cite{yangfisher75}, figure 3.}
\end{figure}
This decomposition is
in agreement with McCoy and Wu's
results\cite{mccoybook,mccoy67b}. Indeed in the thermodynamic limit, we
can use the following identity $(a>b)$,
\bb\nn \frac{1}{2\pi}\int_0^{2\pi}\frac{\dd
\theta}{a+b\cos\theta}=\frac{1}{ \sqrt{(a-b)(a+b)}}, \ee
to obtain
\bb\nn 
\int_0^{2\pi}\frac{\dd \theta_p}{2\pi}
\frac{1}{(1+t^2)^2-2t(1-t^2)(\cos \theta_p
+\cos\theta_{q+\fd})}=\frac{1}{ \sqrt{R(\theta_{q+\fd})}},
\ee
with the function $R$ defined by
\bb\nn
\label{defR}
\fl
R(\theta)=\left[(1+t^2)^2+2t(1-t^2)(1-\cos \theta)
\right]\left[(1+t^2)^2-2t(1-t^2)(1+\cos \theta)\right].  \ee
%
Then the following coefficients can be evaluated in this limit:
\bb \label{resgammalambda}
\gamma^{2D}_{q+\fd}=\frac{2it\sin \theta_{q+\fd}}
{\sqrt{R(\theta_{q+\fd})}},\;\;
\Lambda_{q+\fd}=-\fd+\fd\frac{(1+t^2)(1-2t\cos\theta_{q+\fd}-t^2)}{
\sqrt{R(\theta_{q+\fd})}}.  \ee
Using the previous results and after some algebra, we recover the result 
of McCoy and Wu, and the continuous form for the boundary free energy 
depending on the magnetic field is
\bb
\label{free_field_h_continuous}
-\beta f_{\mathrm{field}}=\frac{1}{4\pi}\int_{-\pi}^{\pi}\dd \theta  \ln
\left(1+\ff{4u^2 t (1+\cos \theta)} {(1+t^2)(1-2t\cos \theta
-t^2)+\sqrt{R(\theta)}}\right).  \ee 
The expression (\ref{free_field_h})
of the free energy allows the numerical computation of the specific
heat, even for $L$ small. The results are presented on the 
figure \ref{Free_vH} for various magnetic fields.\\
\section{Boundary two point correlation function and magnetisation}

In this section, we compute the boundary spin-spin correlation
functions  along the boundary line between two different  sites
$(1,k)$ and $(1,l)$, $k<l$, in the case where the magnetic field is
uniform. This is  the easiest case since we can use simple properties
of the Grassmann magnetic fields.  Using (\ref{pfu}), we have
\bb \left< \sigma_{1k} \sigma_{1l} \right> &\propto
\sump{\sigma}\left[\sigma_{1k}\sigma_{1l}e^{-\beta \cal{H}}\right], \\
\nn &
=\sump{\sigma}\left[\prod_{m,n=1}^{L}\psi^{(1)}_{mn}\psi^{(2)}_{mn}
\cdot\prod_{n=1}^{L}\sigma_{1k}\sigma_{1l}(1+u\sigma_{mn}) \right].
\ee
We then write ($u\neq0$):
\bb\label{trick}
\sigma_{1k}(1+u\sigma_{1k})=u(1+\ff{1}{u}\sigma_{1k}), \ee
and introduce local magnetic fields
$u_n=u+(u^{-1}-u)(\delta_{kn}+\delta_{ln})$ so that
\bb 
\nn \left< \sigma_{1k} \sigma_{1l} \right> \propto
u^2\sump{\sigma}\left[\prod_{m,n=1}^{L}\psi^{(1)}_{mn}\psi^{(2)}_{mn}
\cdot\prod_{n=1}^{L}(1+u_n\sigma_{mn}) \right].  \ee
We remarks that this expression is the PF of the 2D Ising model in the
particular case of an inhomogeneous BMF
(\ref{pfu}).  The integration over the bulk variables is not affected
by this change.  The difference appears \textit{only} in the non local
coupling between the fermionic magnetic fields $H_n$,
\bb \prodd{m=1}{L} e^{u_n H_n}= \exp\left[  \sum_{n=1}^{L}u_n H_n+
\sum_{m=1}^{L-1} \sum_{n=m+1}^{L} u_mu_n H_mH_n \right].  \ee
Using the expression for $u_n$, we obtain
\bb \fl \sum_{m<n}u_mu_nH_mH_n&=
u^2\sum_{m<n}H_mH_n+(1-u^2)(H_kL_k+H_lL_l) +\ff{1-u^2}{u^2}H_kH_l, \\
\nn L_k&=\sum_{n=k}^{L}H_n-\sum_{n=1}^{k-1}H_n, \ee
and therefore,
\bb \nn \fl
\label{c1}
\prodd{n=1}{L} e^{u_n H_n} = \prodd{n=1}{L} e^{u H_n} \cdot
 \left[ 1+ (1-u^2)(H_kL_k+H_lL_l)+ \ff{1-u^2}{u^2}H_kH_l+
(1-u^2)^2)H_kL_kH_lL_l \right].  \ee
Then the two point correlation function can be simply expressed with
correlation functions $\left<H_kH_l\right>$, $\left<H_kL_k\right>$,
and $\left<H_kL_kH_lL_l\right>$
\bb \nn \left<\sigma_{1k}\sigma_{1l}\right> &=&
u^2+u^2(1-u^2)\left<H_kL_k+H_lL_l\right>+(1-u^2)\left<H_kH_l\right>
\\ \nn
&+&u^2(1-u^2)^2\left<H_kL_kH_lL_l\right>.  \ee
The correlation functions involving a product of four Grassmann fields
can be written in term of products  $\left<H_kH_l\right>$  according
to Wick's theorem:
\bb\label{correlations} \left<H_kL_kH_lL_l\right>&=
\left<H_kL_k\right>\left<H_lL_l\right>
-\left<H_kL_l\right>\left<H_lL_k\right>
-\left<H_kH_l\right>\left<L_kL_l\right>.  \ee
Using a Fourier transformation, the two field correlation functions
are expressed, using the usual definitions, as
\bb\label{FourierCorrelation} \left<H_kH_l\right>&=\ff{2i}{L}
\sum_{q=0}^{L/2-1}\left<H_{q+\fd}H_{-q-\fd}\right>
\sin\left[\theta_{q+\fd}(k-l)\right], \\ \nn
&=\ff{2}{L}\sum_{q=0}^{L/2-1}\ff{i\gamma^{2\mathrm{D}}_{q+\fd}}
{\fZ} \sin\left[\theta_{q+\fd}(k-l)\right]. \ee
Each term on the right hand side of equation (\ref{correlations}) can
then be evaluated using the previous result:
\bb
 \left<H_kL_k\right>\left<H_lL_l\right> &=\left (
\ff{1}{L}\sum_{q=1}^{L/2-1}
\ff{i\gamma^{2\mathrm{D}}_{q+\fd}}{\fZ}{\rm \cot}
\ff{\theta_{q+\fd}}{2} \right )^2, \\ 
\nn \left<H_kL_l\right>
&=-\ff{1}{L}\sum_{q=1}^{L/2-1}
\ff{i\gamma^{2\mathrm{D}}_{q+\fd}}{\fZ} \ff{\cos
[\theta_{q+\fd}(k-l+1/2)]}{\sin (\theta_{q+\fd}/2)}, \\
\nn
\left<L_kL_l\right> &= \ff{2}{L}\sum_{q=1}^{L/2-1}
\ff{i\gamma^{2\mathrm{D}}_{q+\fd}}{\fZ} \ff{\sin
[\theta_{q+\fd}(k-l)]}{\sin(\theta_{q+\fd}/2)^2}.  \ee
The magnetisation can be computed the same way. Using the  identity
(\ref{trick}) we obtain
\bb 
\label{magnetizationwithmagneticfield}
\left<\sigma_{1k}\right>= u+u(1-u^2)\left<H_kL_k\right>.  \ee
The connected correlation function is then
\bb \fl \nn \left<\sigma_{1k}\sigma_{1l}\right>-
\left<\sigma_{1k}\right>\left<\sigma_{1l}\right> =
(1-u^2)\left<H_kH_l\right>
-u^2(1-u^2)^2 \left ( \left<H_kL_l\right>\left<H_lL_k\right>
+\left<H_kH_l\right>\left<L_kL_l\right> \right ).  \ee These
correlation functions, and particularly $\left<H_kH_l\right>$, can be
extended for the study of more complex configurations of the boundary
magnetic field.
 Without field ($u=0$), the correlation function between two spins $\sigma_{1k}$
and $\sigma_{1l}$ are simply the correlation function between the 
two Grassmannian fields $H_k$ and $H_l$ and is given by 
equation (\ref{FourierCorrelation}). We can extract from this the dependence 
of the magnetisation per spin $m$ in the thermodynamic limit near
the critical point $T_c$.
Indeed, it is usual to define $m^2$ as the limit of the two point correlation 
function for large separation $r=|k-l|$:
\bb
\lim_{r\rightarrow \infty}\left<\sigma_{1k}\sigma_{1l}\right>=m^2.
\ee
To obtain the main contribution from (\ref{FourierCorrelation})
near $T_c$ in the thermodynamic limit, we use the expressions
(\ref{resgammalambda}) to compute $\fZzero$ in
(\ref{FourierCorrelation}) and then make an 
expansion around $t_c$ in the low temperature limit ($t>t_c$) of the
different quantities.
We first write the correlation function as an integral
\bb\label{correlationintegral}
\left<\sigma_{1k}\sigma_{1l}\right>&=&
\frac{2t}{\pi}\int_0^{\pi}\dd \theta \frac{\sin(\theta)
\sqrt{R(\theta)}}{S(\theta)}\sin(\theta r),
\\ \nn
S(\theta)&=&\frac{1}{4}\left [
\sqrt{R(\theta)}+(1+t^2)(1-2t\cos\theta-t^2)\right
]^2+4t^4\sin^2\theta,
\ee
and expand $R$ and $S$ for $\dt=t-t_c$ and $\theta$ small, 
which is the region where the main contribution of the integral comes 
from. 
For $R$, we find the following expansion:
\bb
\nn
\sqrt{R(\theta)}&=&\dt(1+t_c^2)(1+t_c+\sqrt{2})\left [
1+\frac{t_c(1-t_c^2)}{2(1+t_c+\sqrt{2})}\tth^2+\ldots \right ],
\\ \nn
&=&\dt(R_0+R_2\tth^2+\ldots)
\ee
where we defined $\tth\dt=\theta$, $R_0=(1+t_c^2)(1+t_c+\sqrt{2})$ and
$R_2=t_c(1-t_c^4)/2$. For $S$, we obtain:
\bb
\nn
S(\theta)=\dt^2\left( S_0+S_2\tth^2+\ldots \right ),
\ee
where $S_0$ and $S_2$ are numerical coefficients evaluated at $t_c$:
$S_0=16(3-2\sqrt{2})\simeq 2.745$,
$S_2=4t_c^4\simeq .118$.
After some algebra,
we obtain the following behaviour near $t_c$ of the two point correlation 
function:
\bb
\left<\sigma_{1k}\sigma_{1l}\right>\simeq \frac{2t\dt}{\pi}
\int_{0}^{\infty}\dd x \frac{xR_0
}{S_0(r\dt)^2+S_2x^2}\sin x.
\ee
For $r\dt$ small enough, the integral is a constant $\pi R_0/2S_2$, and $m$ 
is then proportional to $\sqrt{\dt}$, which gives the mean field 
exponent $\beta=1/2$ for the boundary magnetisation.
 With the presence of a small BMF, at $T_c$, we use 
equation (\ref{magnetizationwithmagneticfield}) to compute directly
the magnetisation:
\bb\label{integralMwithMF}
\left<\sigma_{1k}\right>\simeq u\left<H_kL_k\right>
=\frac{ut}{\pi}\int_{0}^{\pi}\dd \theta\frac{(1+\cos \theta )\sqrt{R(\theta)}}
{S(\theta)+2tu^2(1+\cos \theta)\sqrt{R(\theta)}}.
\ee
The quantities $R$ and $S$ are easily evaluated at $t_c$:
\bb\nn
R(\theta)=16C_0(1-\cos\theta)(3-\cos\theta),
\\ \nn
S(\theta)=8C_0(1-\cos\theta)
\left [
3-\cos\theta+
\sqrt{2(1-\cos\theta)(3-\cos\theta)}\right ],
\ee
with $C_0=17-12\sqrt{2}$. When $u$ is small, the main contribution
in (\ref{integralMwithMF}) is when the denominator is small, or 
when $\theta$ is close to zero. In this case we have $\sqrt{R(\theta)}\simeq
4\sqrt{C_0}\theta$ and $S(\theta)\simeq 8C_0\theta^2$. Replacing these
values in (\ref{integralMwithMF}) we obtain in the limit where $u\ll 1$:
\bb
\left<\sigma_{1k}\right>
\simeq
\frac{ut}{\pi}\int_{0}^{\pi}\dd \theta\frac{1}{
\sqrt{C_0}\theta+2tu^2}\simeq -\frac{2t}{\pi\sqrt{C_0}}u\log u.
\ee
These results agree with McCoy and Wu's paper \cite{mccoy68} and 
we can extend the method to the case
of an inhomogeneous BMF, as we will see in the
next section.%
\section{An example of inhomogeneous boundary magnetic field}
The previous solution in the presence of a  
uniform BMF (configuration $C_a$) on one or two 
sides has be presented in 
different publications
(see for example \cite{yangfisher80}, \cite{puzzo02} and \cite{bugrij90}), but
the existence of a 1D Gaussian action allows us to solve more general 
configurations of the BMF.
In this section, we illustrate that considering the simplest case of a non 
uniform field at the boundary:
Between sites $(1,1)$ and $(1,l)$ we impose a field $h$, and  between
sites $(1,l+1)$ and $(1,L)$ a field $-h$ (configuration noted $C_b$).
This problem could be interesting for the study of interfaces
\cite{abraham80,abraham82,abraham88,forgacs88,maciolek96,ebner90}
and to our knowledge it has not been solved exactly for arbitrary values
of $h$ on a finite lattice.\\
Using our method, it appears that the evaluation of the PF $\pf(h;l)$ is a
simple extension of the previous calculation. Indeed, the product 
(\ref{c1}) depending on the field can be simply expressed as
\bb\nn \exp \left (\sum_{m<n}u_mu_nH_mH_n\right )= \exp \left
(u^2\sum_{m<n}H_mH_n-2u^2\sum_{m=1}^{l}H_m\sum_{n=l+1}^{L}H_n \right
).  \ee
Then we have
\bb \eqalign{
\label{Qfh}
\pf(h;l)&=\pf_0  \mathrm{Tr}\left[e^{\action_{1D}}
\left(1-2u^2\sum_{m=1}^{l}\sum_{n=l+1}^{L}H_mH_n\right) \right],\\ \nn
&= \pf(h)\left(1-2u^2\sum_{m=1}^{l}\sum_{n=l+1}^{L}\left<H_mH_n\right>_{\action_{1\mathrm{D}}}\right),
} \ee
since the other terms from the exponential expansion all include  the
square of linear Grassmann sums and are therefore zero.  

Using the Fourier transformation (\ref{FourierCorrelation}), we obtain :
\bb \frac{\pf(h;l)}{\pf(h)}=1+\frac{4u^2}{L}
\sum_{q=0}^{L/2-1}\frac{i\gamma^{2\mathrm{D}}_{q+\fd}}{\fZ}
\frac{\sin \theta_{q+\fd}l}{1-\cos \theta_{q+\fd}}.  \ee
The free energy difference between the 2 field configurations $C_b$ 
and $C_a$ is positive and is equal to

\bb 
\label{fint2D1} 
-\beta f_{\mathrm{int}}=\ln\left(1+4u^2\frac{1}{L}
  \sum_{q=0}^{L/2-1}\frac{i\gamma^{2\mathrm{D}}_{q+\fd}}{\fZ}\frac{\sin \theta_{q+\fd}l}{1-\cos\theta_{q+\fd}} \right). 
\ee 
This term is added to the total free energy (\ref{Ftotal}) and for $l=L/2$,
${\mathrm{sin}}(\theta_{q+\fd}l)$ is simply equal to $(-1)^q$. If the 
transverse direction has an infinite size, we can use the expressions
(\ref{resgammalambda}) to obtain
\bb
\label{fint2D2}
-\beta f_{\mathrm{int}}=\ln\left(
1-8u^2\frac{1}{L}
\sum_{q=0}^{L/2-1}
\frac{(-1)^q\cot(\theta_{q+\fd}/2)}{\sqrt{R(\theta_{q+\fd})}\fZ}
 \right), 
\ee 
or more explicitly
\bb \nn
\fl
-\beta f_{\mathrm{int}}=\ln \left(
1-\ff{16tu^2}{L}
\sum_{q=0}^{L/2-1}
\ff{(-1)^q\;\cot(\theta_{q+\fd}/2)}{(1+t^2)(1-2t\cos\theta_{q+\fd}-t^2)+
4tu^2(1+\cos \theta_{q+\fd})+\sqrt{R(\theta_{q+\fd})}}
\right).
\ee
In the 1D Ising model, for a system of $L$ spins with periodic boundary 
conditions
and with a field configuration $C_b$ identical to the 2D boundary line
with $l=L/2$, a similar result can be obtained, either by the
transfer matrix or the Grassmannian methods of the section 
\ref{1Dsection}.  
 In the first case, we obtain

\begin{figure}[!h]
\begin{center}
\psfrag{fint/L}{$f_\mathrm{int}/L$}
\psfrag{h/J}{$h/J$}
\includegraphics[scale=0.45,angle=0]{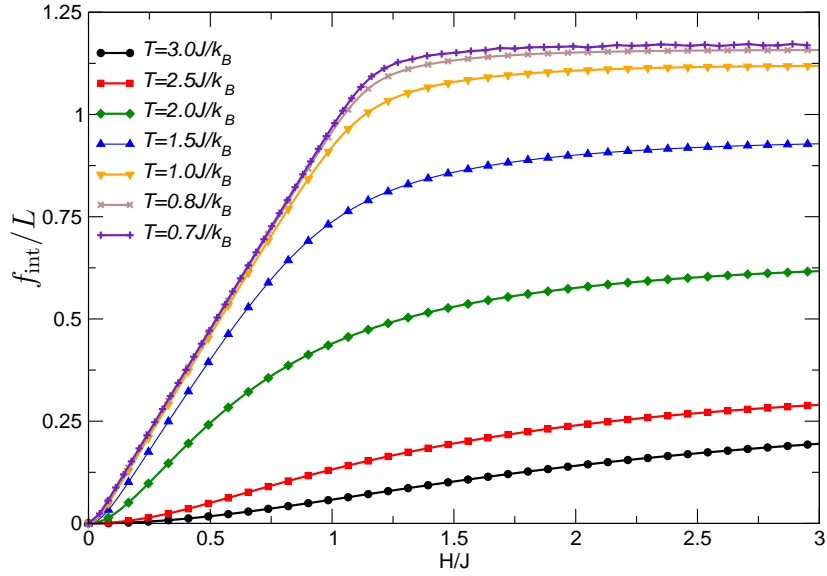}
\caption{Boundary free energy $f_{\mathrm{int}}/L$ as function of $h$ 
for $L=20$, $h_c=1.2$, for various values of the temperature. 
Notice the transition at $h=h_c$ corresponding to the reversing of half
the boundary spins.}
\label{interfaceH}
\end{center}
\end{figure}

\begin{figure}[!h]
\begin{center}
\psfrag{fint/L}{$f_\mathrm{int}/L$}
\psfrag{t=tanh(J/kbT)}{$t=\tanh(J/k_BT)$}
\includegraphics[scale=0.45,angle=0]{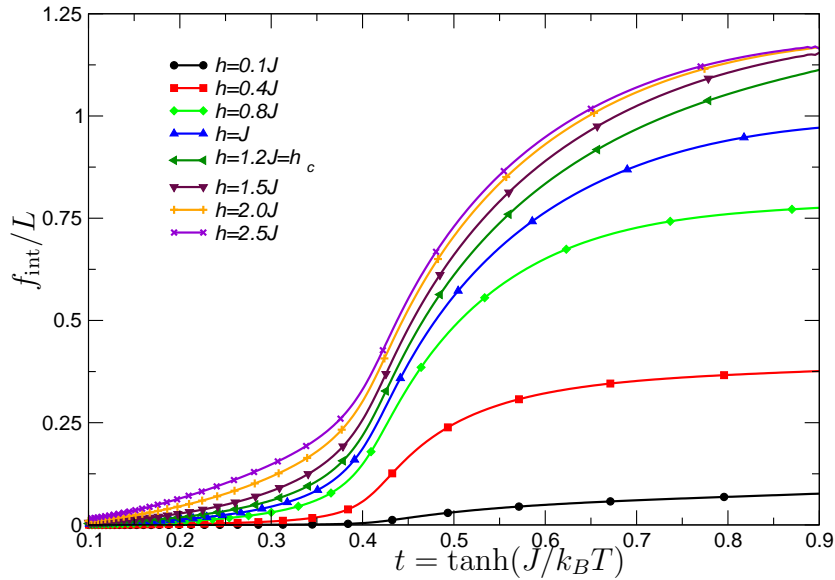}
\caption{Boundary free energy $f_{\mathrm{int}}/L$ for $L=20$, $h_c=1.2$,
as a function of $\tanh(J/k_BT)$ for various values of the magnetic 
field $h/J$. 
}
\label{interfaceT}
\end{center}
\end{figure}

\bb\nn
\fl-\beta f_{\mathrm{int}}^{1\mathrm{D}}=\ln \left (
(1-t)^2+
\frac{8tu^2[4t(1-u^2)]^{L/2}}{
(1+t-\sqrt{(1-t)^2+4tu^2})^{L}+(1+t-\sqrt{(1-t)^2+4tu^2})^{L}} \right )
\\ 
\label{fint1DL}
-\ln [(1-t)^2+4tu^2]
\ee
and in the thermodynamic limit this leads to
\bb\label{fint1D}
-\beta f_{\mathrm{int}}^{1\mathrm{D}}=\ln \left [
\frac{(1-t^2)^2}{(1-t)^2+4tu^2} \right ].
\ee
With the Grassmannian fields, we obtain a different expression but rigorously
identical to equation (\ref{fint1DL}):
\bb\label{fint1DSum}
\fl
-\beta f_{\mathrm{int}}^{1\mathrm{D}}=\ln \left (
1-\frac{8tu^2}{L}\sum_{q=0}^{L/2-1}
\frac{(-1)^q\cot(\theta_{q+\fd}/2)}
{1+t^2-2t\cos(\theta_{q+\fd})+4tu^2\cos(\theta_{q+\fd}/2)^2}
\right ).
\ee
In the thermodynamic limit and at zero temperature, $f_{\mathrm{int}}^{1\mathrm{D}}$
is equal to $4J$ for any non zero value $h$, which is the energy difference 
between the 2 ground states of $C_\mathrm{b}$ and $C_\mathrm{a}$, all spins following 
the magnetic field direction in both cases.
 In the 2D case however, if we suppose that the size is infinite in the
transverse direction, for small values of the magnetic field the 
boundary spins point all in the same direction (imposed by the bulk spins, up
for example) below a critical value
of the field $h_c=J(1+4/L)$ and for the $C_\mathrm{b}$ configuration. 
Indeed the interaction between boundary
spins and the ones in the bulk are strong enough that a small field $-h$
is not sufficient to reverse these spins.
In this case $f_{\mathrm{int}}=hL$ and is therefore
extensive contrary to the 1D case. Above $h_c$, the field $-h$ is strong 
enough to reverse half of the spins that were up and therefore
$f_{\mathrm{int}}=h_cL$ due 
to the frustrated couplings with the neighbouring spins in the bulk. 
Figure \ref{interfaceH} represents the free energy $f_{\mathrm{int}}/L$
as a function of the BMF for different values of the 
temperature. $L=20$ and $h_c=1.2J$. There is a change of the curve slope 
at $h=h_c$ and low temperature curves suggest the previous reversal picture. 
%
%
%
Figure \ref{interfaceT} represents the contribution $f_{\mathrm{int}}/L$ 
as a function of temperature for various values of the magnetic fields.
For $h>h_c$ the curves saturate at zero temperature to the value
$f_{\mathrm{int}}/L=h_c$ as expected. 
%
%
%
To check the low temperature behaviour of $f_{\mathrm{int}}$, we can perform 
an expansion of equation (\ref{fint2D2}) for $h\ll 1$ and $T\ll 1$. 
In this case, we assume that $t\simeq 1$ and $u\simeq 1-2\exp(-2h/k_BT)$. 
We obtain
\bb
-\beta f_{\mathrm{int}} \simeq \ln\left(
1-4u^2\frac{1}{L}\sum_{q=0}^{L/2-1}
\frac{(-1)^q
\cot(\theta_{q+\fd}/2)}{(1+u^2)-(1-u^2)\cos\theta_{q+\fd}}
\right).
\ee
The sum inside the logarithm can be computed using equations (\ref{fint1DL})
and (\ref{fint1DSum}), with $t=1$. We obtain
\bb
-\beta f_{\mathrm{int}}=\ln\left(
\frac{2(1-u^2)^{L/2}}{(1-u)^L+(1+u)^L}
\right )\simeq -\beta hL
\ee
These results allow us also to study quite precisely the effect of an
inhomogeneous magnetic field on the spins inside the bulk for a finite
transverse size system or a fixed ratio between the sizes of the two directions
(square or rectangular system). 
In this case, the previous domain wall that appears for $h>h_c$ due to the
reversal of half of the boundary spins can propagate or jump inside 
the bulk for 
sufficiently high temperature in order to lower the free energy. This 
may cause the total magnetisation to cancel if this wall separates two regions 
of equal number of opposite spins.
An exhaustive study will be published in a forthcoming paper.

\section{Conclusion}
In this article we have presented a generalisation of the Plechko's
method to the 2D Ising model in the case of a inhomogeneous BMF. 
We showed that for any configuration of the BMF, 
this model can be mapped onto a 1D model with
a Gaussian Grassmannian action (\ref{res2D1Dgen})
similar to the simpler case of the 1D Ising in a magnetic field. 
Results have been obtained for the
homogeneous case, in order to validate our results with the ones obtained
by McCoy and Wu\cite{mccoy67b}, and an extension is made to a special
inhomogeneous case where an interface develops from one side of the 
boundary. 
A simple expression of the boundary partition function in term
of a 1D action is given. Contrary to the 1D case, the interface free 
energy appears to be extensive, proportional to the system size $L$.
A simple argument gives a zero temperature critical field $h_c$ to be 
the field above 
which a domain wall appears on the boundary and may eventually propagate inside
the bulk at higher temperature. 
 This generalisation of Plechko's method for the Ising model with non uniform 
BMF appears to be simpler than methods based on 
transfer matrix theories. Further developments are possible, for example 
in the study of wetting by a defect plane \cite{ebner90}. More generally,
it appears however that this method does not answer the question of 
the 2D Ising model with an homogeneous magnetic field in the bulk
\cite{zamolo1,zamolo2,mussardoMFTc,delfinoMFTc}, even for the 3D Ising 
problem, since the operations 
using mirrors symmetries generate a Grassmannian action which is no longer
quadratic. This is due to the fact, in the 2D case, that linear Grassmann 
terms proportional to the field do not commute each other easily. 
 The mapping of the boundary region onto a 1D action could be used to 
study precisely the properties of boundary random magnetic fields.
For the 2D Ising case, these random fields are found to
be marginal \cite{Cardy91} and logarithm corrections to the pure
case have been checked numerically \cite{rfim}. Analytical computations 
based on the 1D general action (\ref{res2D1Dgen}) can be performed using 
possibly random matrix theory (see also \cite{mussardoRMF} in the 
context of the Conformal Field Theory).

\ack We acknowledge Francois Delduc, Peter C.W. Holdsworth and  Jean
Richert for useful discussions.

\bibliographystyle{unsrt} 


\end{document}